\documentclass[11pt,english,american]{article}
\usepackage{amsmath}
\usepackage{ amssymb }

\usepackage{ae,aecompl}
\usepackage[T1]{fontenc}
\usepackage[latin9]{inputenc}
\usepackage[a4paper]{geometry}
\usepackage{textcomp}
\usepackage{amstext}
\usepackage{graphicx}
\usepackage{cite}
\usepackage{color}

\usepackage{authblk}
\usepackage{babel}

\newcommand{\arcsec}{\ensuremath{^{\prime\prime}}}

\begin{document}

\title{Shell galaxies as laboratories for testing MOND}
\author[1,2]{M. B\'{i}lek}
\author[1,3]{I. Ebrov\'{a}}
\author[1,2]{B. Jungwiert}
\author[4]{L. J\'{i}lkov\'{a}}
\author[1,5]{K. Barto\v{s}kov\'{a}}

\affil[1]{Astronomical Institute, Academy of Sciences of the Czech Republic, Bo\v{c}n\'{i} II 1401/1a, CZ-141\,00 Prague, Czech Republic\\
       bilek@asu.cas.cz}
\affil[2]{Faculty of Mathematics and Physics, Charles University in Prague, Ke~Karlovu 3, CZ-121 16 Prague, Czech Republic}
\affil[3]{Institute of Physics, Academy of Sciences of the Czech Republic, Na~Slovance 
1999/2 CZ-182\,21 Prague, Czech Republic}
\affil[4]{ Leiden Observatory, Leiden University, P.O.~Box~9513, NL-2300RA Leiden, The~Netherlands }
\affil[5]{Department of Theoretical Physics and Astrophysics, Faculty of Science, Masaryk University, Kotl\'a\v{r}sk\'a 2, CZ-611\,37 Brno, Czech Republic}
\renewcommand\Authands{ and }

\maketitle

\begin{abstract}
\noindent
Tests of MOND in ellipticals are relatively rare because these galaxies
often lack kinematic tracers in the regions where the MOND
effects are significant. Stellar shells observed in many elliptical
galaxies offer a promising way to constrain their gravitational field.
Shells appear as glowing arcs around their host galaxy. They are
observed up to $\sim$100\,kpc. The stars in axially symmetric shell systems move
in nearly radial orbits. The radial distributions of shell locations and the spectra of stars in shells can
be used to constrain the gravitational potential of their host galaxy. The symmetrical shell systems, being especially suitable for these studies, occur in approximately 3\% of all early-type galaxies.  Hence the shells substantially increase the number of ellipticals in which MOND can be tested up to large radii. In this paper, we review our work on shell galaxies in MOND. We summarize the paper B\'{i}lek et~al.~(Astron. Astrophys., 559, A110, 2013), where we demonstrated the consistency of shell radii in an elliptical NGC\,3923 with MOND, and the work B\'{i}lek et~al.~(2014, arXiv:1404.1109), in which
we predicted a giant ($\sim$200 kpc), yet undiscovered shell of NGC\,3923. We
explain the shell identification method, which was used in these two
papers. We further describe the expected shape of line profiles in shell
spectra in MOND which is  very special due to the direct relation
of the gravitational field and baryonic matter distribution (B\'{i}lek et~al., 2014, in preparation).
\end{abstract}

\section{Introduction} \label{sec:intro}
The Modified Newtonian dynamics (MOND) and its implications have been successfully tested in all types of disk and dwarf galaxies and its numerous aspects have been even verified in interacting galaxies \cite{famaey12}.
However, MOND is difficult to be tested in giant ellipticals, up to the large radii, where MOND predicts large mass discrepancies, due to the lack of kinematics tracers up in these regions. Apart from a~few exceptions, there are no objects in known orbits, similar to the gas clouds in spiral galaxies, which would enable a direct measurement of the gravitational acceleration. Jeans analysis of dynamics of stellar or planetary nebulae can be used to measure the gravitational field, but its results are ambiguous since the anisotropic parameter is unknown. Hot X-ray emitting coronal gas is present mostly in the central clusters galaxies, which may be influenced by their neighbors.   Finally, gravitational lensing is not able to probe gravitational field in the low-acceleration regime (see Sect.~1 in Ref.~\cite{milg12} for more details and a~recent review of tests of MOND in ellipticals).

Stellar shells, observed in many ellipticals, offer an interesting alternative to the established methods mentioned above. Two ways to constrain the gravitational potential in ellipticals with shells were described -- using the radial locations of shells and spectra of stars in shells.  As we explain in Sect.~\ref{sec:rad}, the radius of a~shell in an~axially symmetric, so-called Type\,I, shell system is a~function only of the shape of the galactic gravitational potential and the time elapsed since the merger.  Hence, the observed radii of shells can be compared with the prediction of given gravitational potential. This is a~very interesting way for testing MOND, which predicts the dynamics only on the basis of the distribution of the baryonic matter. The three-dimensional shape of the galaxy, needed to built the MOND potential, can be, for a shell elliptical, constrained from the morphology of its shells \cite{DC86}. Moreover, shells often extend very far from the center of their host galaxy. The elliptical NGC\,3923 (Fig.~\ref{fig:N3923}), which is the main subject of this paper, possess the biggest known shell in the Universe. Its radius is more than 100\,kpc (Table~\ref{tab:shpos}) and, according to the MONDian model of the galaxy NGC\,3923 from B\'{ i}lek et~al. \cite{bil13}, it is exposed to gravitational acceleration of $a_0/5$ (Table~\ref{tab:shpos}). We predicted theoretically the presence of a~new --  yet undiscovered -- shell at the distance of about 220\,kpc \cite{bil14a}, which would extend down to the acceleration of $a_0/10$. It was possible to test MOND in such a low acceleration only in two ellipticals so far -- in NGC\,720 and NGC\,1521 \cite{milg12}. 

Compared to measuring the radial positions of shells, constraining the
potential of a shell galaxy form the spectral line profiles is more
instrumentally demanding. No observations of shell spectra have been
presented so far. However, the fast developing observational facilities
will make this method possible to apply in the near future. The spectral line of a shell is quadruple-peaked \cite{ebrova12}. The separations of the peaks is directly related to the gravitational acceleration at the radius of the shell edge and the phase velocity of the shell.  However, their separation in MOND can be predicted theoretically  \cite{bil14b}.
 
Type\,I shell galaxies are interesting from the point of view of MOND also for another reason. So far, MOND has been tested only in systems the constituents of which move in nearly circular orbits (like disk galaxies,  \cite{thingsmond, rotcurv1, rotcurv2}), randomly distributed orbits (dwarf galaxies,  \cite{anddwarf}, elliptical galaxies, \cite{sanders2010}) and ellipse-like orbits (polar rings, \cite{lughausen13}). To our knowledge, MOND has never been tested for particles in radial trajectories. Since the inspiration for MOND originally came from the flatness of rotational curves of disk galaxies \cite{milg83a}, it is important to test MOND also for strongly non-circular orbits. It is not known whether MOND should be interpreted as a~modified gravity theory or as a~modification of the law of inertia (or as a~combination of both). In the first case, the kinematic acceleration of a~test particle depends only on the vector of gravitational acceleration at the immediate position of the particle; in the second, the kinematic acceleration can generally depend on the whole trajectory of the particle since the beginning of the Universe. In the case of modified inertia, the MOND algebraic relation $a\mu(a/a_0) = a_N$, deduced for particles in circular orbits, does not have to be necessarily valid for particles oscillating along a~line. Furthermore, in many cases the stars constituting the shells periodically travel between the Newtonian ($a\gg a_0$) and deep-MOND ($a\ll a_0$) region of their host galaxy, unlike the stars in disk galaxies, which keep to be exposed to the gravitational field of nearly constant magnitude.

In Sect.~\ref{sec:shgal}, we give a brief summary of general characteristics of shell galaxies. The  shells kinematics is explained in Sect.~\ref{sec:rad}. In Sect.~\ref{sec:kin}, we summarize the method of potential constraining form shell spectra. We will discuss the expected line profile for the case of MOND in Sect.~\ref{sec:shvel}. In Sect.~\ref{sec:shid}  we describe the method for constraining the gravitational potential from the shell radial distribution and  we present our results achieved when the method is applied on  NGC\,3923: the consistency of shell positions with MOND and the prediction of a new shell. Finally, we summarize our paper in Sect.~\ref{sec:sum}.

\section{Shell galaxies}
\label{sec:shgal}
Shells, also called ripples, were first noticed
by Halton Arp in his \textit{Atlas of Peculiar Galaxies} \cite{1966apg..book.....A, 1966ApJS...14....1A}. 
The only specialized list of shell galaxies is \textit{A catalogue
of elliptical galaxies with shells} of Malin and Carter \cite{malin83}, which is however only partial.
The authors presented a catalogue of 137 galaxies that exhibit shell or ripple
features at large distances from the galaxy or in the outer envelope. 

\subsection{Characteristics of shells}

Shell galaxies are mostly ellipticals containing fine stellar structures
in the form of open, concentric arcs that do not cross each other. In
general, they tend to have sharp outer boundaries, but many of them
are faint and diffuse. Shells exhibit a low surface brightness from
about 24\,mag$/$arcsec\textsuperscript{2} (in the V band) up to the current detection limit (about 30\,mag$/$arcsec\textsuperscript{2}). 
The number of shells in a galaxy ranges from 1 to 27 (for the richest known shell galaxy NGC\,3923, \cite{prieur88,sikkema07}), but a large fraction of  Malin and Carter's catalog (reaching, however, only to 26.5\,mag$/$arcsec\textsuperscript{2} in the B band) consists of galaxies with less than 4 shells. Shells are of  stellar nature, but H\,I \cite{1994ApJ...423L.101S,1995ApJ...444L..77S,1997AAS...191.8212P,1997ASPC..116..362S,2001AJ....122.1758B} and molecular gas \cite{2000A&A...356L...1C,2001A&A...376..837H} associated with shells were detected in some cases.

Three different morphological types of shell galaxies were defined \cite{prieur90, wilkinson87}: Type\,I 
-- axisymmetric shell systems confined into a double cone, the shells are interleaved in radius (i.e.,
the next outermost shell is usually on the opposite side of the nucleus), their
separation increases with radius, and their symmetry axis is well-aligned with the major axis of the galaxy (e.g., NGC\,3923, Fig.~\ref{fig:N3923}); Type\,II -- randomly distributed arcs all
around a rather circular galaxy; Type\,III -- a more complex
structure or too few shells to be classified.  These three types occur in approximately the same fraction \cite{prieur90}.

\subsection{Occurrence of shell galaxies}

Shells were originally observed in early-type galaxies (E,
E/S0 or S0) but also some late-type galaxies with shells were
reported later \cite{schweisei88,2010AJ....140..962M}. The tidal
structure including shells in M\,31 is already well studied
\cite{fardal07,fardal12}. There are even indications that a shell can be present in our Galaxy \cite{helmi03,2013ApJ...766...24D}.
Shell structures were reported in
the Fornax dwarf spheroidal galaxy \cite{coleman04,2005AJ....129.1443C}.

Shell galaxies occur in 6\% of lenticular galaxies, 10\% of ellipticals, and
around 1\% of spirals \cite{schweisei88} (see also \cite{schweizer83, malin83,2011MNRAS.410.1550R,atkinson13}).
In a complete sample of 55 nearby luminous
elliptical galaxies (detection limit 27.7\,mag$/$arcsec$^2$ in V band), at least 22\% of galaxies have shells, making them the most common interaction signature \cite{tal09}. 

The occurrence of shell galaxies turns to be strongly environmentally
dependent. In Malin and~Carter's catalog,
47.5\% shell galaxies are isolated, 30.9\% occur in loose groups,
only 3.6\% occur in clusters or rich groups, 18\% occur in groups
of two to five galaxies \cite{malin83}. 

However, the true abundance of shell galaxies can still be different
from what has been summarized here. It crucially  
depends on which galaxies are classified as shell galaxies and on the ability to detect
faint shells in otherwise plain-looking galaxies. It was 
suggested that the majority of tidal features in early-type galaxies
occur at the surface brightness near 28\,mag$/$arcsec\textsuperscript{2} or fainter \cite{atkinson13}. 
This means that the abundance of shell galaxies could increase in
future with better instrumental equipment. 

\subsection{Theories of origin}

\begin{figure}
\begin{centering}
\includegraphics[width=\textwidth]{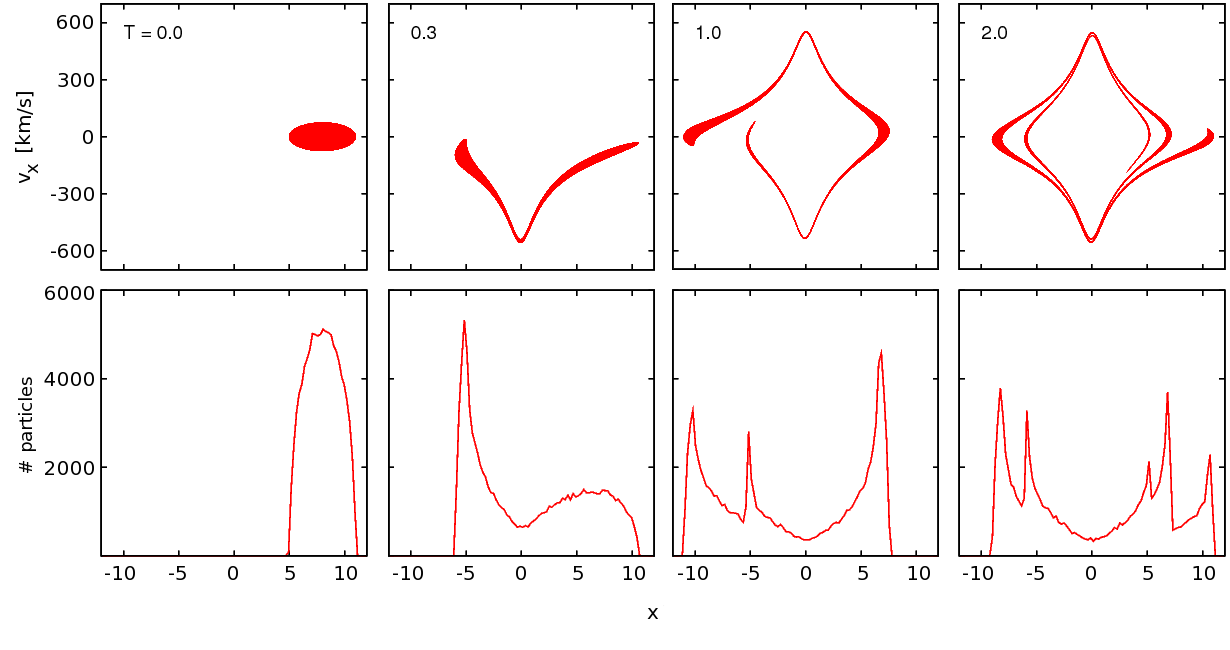}
\end{centering}
\caption{ One-dimensional demonstration of phase wrapping. The picture shows the time evolution of a cloud of test particles falling into the Plummer potential in the $v-x$ space (upper row) and the particle radial density
(bottom row). The $x$ axis is centered with the center of the potential.}
\label{obr.phase-wrap} 
\end{figure}

Several theories of an internal origin of shells were proposed, involving
bursts of star formation in a shocked galactic wind \cite{1980Natur.287..613F,1985ApJS...58...39B,1985ApJ...291...80W},
a modest outburst causing a period of star-formation in an outward-moving
disturbance \cite{1987MNRAS.229..129L}, or a hot supernova-driven
galactic wind \cite{1987ApJ...319..601U}. However, all these scenarios were shown improbable by 
the later observations of color indices of shells
\cite{1990AJ....100.1073M,1999MNRAS.307..967T,sikkema07,2013ApJ...773...34G},
which exclude the presence of a significant fraction of young stars
in shells. Shell colors are rather similar to the color of the host
galaxy, but they could be slightly redder and some (especially the outer)
shells are slightly bluer. If the shells were formed in a burst of
star formation long time ago, they would not survive to the present
days with the sharpness they exhibit \cite{prieur90}.

A more viable theory of the shell origin is the Weak
Interaction Model \cite{wim90,wim91}.
In this model, shells are density waves in a thick disk
population of dynamically cold stars induced by a weak interaction with a
smaller galaxy which had flown near the host galaxy. 
However, there is evidence against this model at least in some shell galaxies: 
an inconsistency of the surface brightness and colors of shells and their host galaxy \cite{fort86,prieur88,1999MNRAS.307..967T,sikkema07}, 
a minor axis rotation \cite{minrot}, and no indication of the thick disk \cite{wilkinson00}.

\begin{figure}
\begin{centering}
\includegraphics[width=\textwidth]{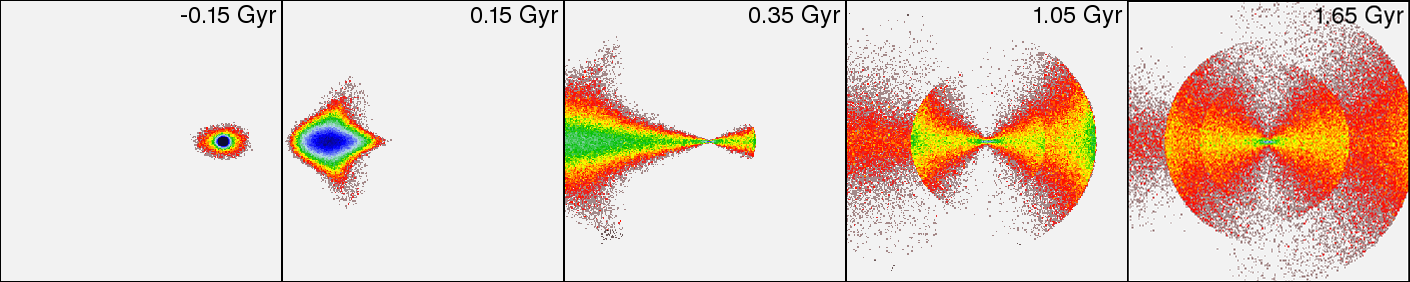}
\par\end{centering}

\caption{%
Snapshots from the Newtonian simulation of a radial minor merger. Only the surface density of particles
originally belonging to the secondary is displayed. Panels, 300$\times$300\,kpc, are centred on the center of the primary with the total mass of 8.2$\times10^{12}$\,M$_{\odot}$.
\label{obr.phase-wrap-3D} 
}
\end{figure}

The most successful scenario is the model of a radial minor merger of Quinn \cite{quinn84}.
When a small galaxy (secondary) enters the scope of influence of a
big elliptical galaxy (primary or host) on a roughly radial 
trajectory, it splits up and its stars begin to oscillate in the potential
of the big galaxy which itself remains mostly unaffected. The stars have 
the slowest speed in their apocenters, and thus tend to
spend most of the time there. They pile up and producing arc-like structures
in the luminosity profile of the host galaxy. This mechanism is also
known as phase wrapping due to its typical pattern in the
space of galactocentric distance and radial velocity. It is illustrated on the one dimensional example in Fig.~\ref{obr.phase-wrap}. 
Results of a three-dimensional simulation are displayed in Fig.~\ref{obr.phase-wrap-3D}.
The shell edges are technically caustics of the mapping of
the phase-space density into the physical space \cite{1989ApJ...346..690N}.
In this model, shells are naturally interleaved in radius, aligned
with the merger axis and their separation increases with radius.

The formation of shell systems was successfully simulated by the accretion of a disk \cite{quinn84} as well as an elliptical  secondary \cite{DC86, hq88}. From the sharpness of observed shells and their
luminosity fraction of the total luminosity of a host galaxy, it
can be judged that the original mass ratio of progenitor galaxies was
from {\small $\sim$}1:100 to {\small $\sim$}1:10. Nevertheless 
 shells have been produced in a simulation of a merger of two elliptical galaxies with the
mass ratio of 1:2 \cite{2005MNRAS.361.1030G,2005MNRAS.361.1043G}.  Hernquist \&~Spergel were even able to
create shells in a simulation of a merger of two identical disk galaxies \cite{majorm}.
On the other hand, Gonz\'{ a}lez-Garc\'{ i}a~\&~Balcells \cite{2005MNRAS.357..753G} noticed the lack of
shells in the remnants of simulated mergers of equal-mass disks. They
concluded that the perfect alignment of the disk rotation axes with the orbital
angular momentum may have favored the formation of shells in the model
of Hernquist \&~Spergel.

\section{Radial distribution of shells}
\label{sec:rad}
In the radial minor merger model, the shells are density waves receding from the host galaxy
center. They are composed of different stars in every moment. The positions of
the shell edges are connected to the potential of the host
galaxy (the primary). 

We approximate the shell system with a simplified model, in which all the stars belonging to the shells were released at the moment, when the secondary reached the center of the primary. The primary is spherically symmetric. The stars are in radial trajectories, their velocities are isotropically distributed with magnitudes smaller than the escape velocity.

The shells occur close to the radii where  stars are located
in their apocenters in the given moment. Such radii $r_{\mathrm{A,}n}$
must satisfy the relation \cite{quinn84, hq87}
\begin{equation}
t=(n+1/2)P(r_{\mathrm{A,}n}),
\label{eq:pos}
\end{equation}
where $t$ is the time elapsed since the stars were released. 
The shell number $n$ ($n=0,1,2\ldots$) corresponds to the number of oscillations 
(from one apocenter to the following one)
that the stars at the shell edge are about to complete. 
The period of radial motion $P(r)$ at a galactocentric radius $r$ in
the host galaxy potential $\phi(r)$ reads 
\begin{equation}
P(r)=\sqrt{2}\int_{0}^{r}\left[\phi(r)-\phi(r')\right]^{-1/2}\mathrm{d}r'.
\label{eq:per}
\end{equation}
Actually, due to the nonzero phase velocity of shells, the stars lie on the shell edge shorty before they reach their apocenters \cite{DC86}. 

The shell moves approximately with the phase velocity of the radius $r_{\mathrm{A,}n}$, which can be obtained by the derivative of Eq.~(\ref{eq:pos}) with respect to time \cite{quinn84}
\begin{equation}
v_{\mathrm{s},n}=\frac{1}{n+1/2}\left(\mathrm{d}P(r)/\mathrm{d}r\right)^{-1}.
\label{eq:dpdr}
\end{equation}
We can clearly see from this relation that any further turning point (turning
point with higher $n$) at the same radius moves more slowly than
the former one. 

The exact positions of a shell in this simplified model cannot be easily expressed. 
They are given by a solution of an integral equation which does not have a known analytical solution (equation (5) of Ref.~\cite{ebrova12}).
The approximation for the radius of a shell given by Eq.~(\ref{eq:pos}) can differ from the precise value by more than 10\% \cite{ebrova12}. A better modeling of the shell radius can be achieved by using the corrected shell position $r_{\mathrm{s},n}$
\begin{equation}
r_{\mathrm{s},n}=\phi^{-1}\left[\phi\left(r_{\mathrm{A,}n}\right)-\frac{1}{2}v_{\mathrm{s},n}^{2}\right],
\label{eq:pos2}
\end{equation}
where $\phi^{-1}$ is the inverse function of $\phi$ \cite{bil13}. This correction follows from the conservation of energy of oscillating stars.

It was discovered early \cite{quinn84} using Eq.~(\ref{eq:pos}) that the radial distribution of shells derived from the Newtonian
potential inferred from the observed luminous matter distribution
does not agree with the observed reality. It was found that the observed shape of the shell distribution can be reproduced better by adding an extensive dark matter halo in the primary \cite{DC86, hq87}. The first attempt to test MOND using the shell distribution was made by Hernquist \&~Quinn \cite{hqmond}.  They concluded that the distribution of shells in NGC\,3923 was inconsistent with MOND, but Milgrom \cite{milgsh}
found several substantial deficiencies in their analysis. Finally, Dupraz \&~Combes
\cite{dupcomb87} synthesized successfully a radial distribution
of shells similar to NGC\,3923 without the need of invoking the dark matter. 

They discovered that the multi-generation shell formation and dynamical friction play a major role in the formation of a shell system. One generation of shells is defined as a set of shells that are made of stars released during the same passage
of the secondary through the primary galaxy. If the core of the secondary
galaxy survives the first impact, it can come back, and then the
system would consist of several generations of shells.

If the shell system was created in $N_{\mathrm{max}}$ generations and the age of the $N$-th generation is $t_N$, the set of radii of all shells can be expressed as
\begin{equation}
	\left\{o_I(-1)^{n+1}(-1)^{N+1}r_{\mathrm{s},n}(t_N); n = n_{\mathrm{min},N}, n_{\mathrm{min},N}+1, \ldots, n_{\mathrm{max},N}; N = 1, 2, \ldots,N_{\mathrm{max}} \right\}.
	\label{eq:rall}
\end{equation}
The second and third term express the facts that the shells in a Type~I shell galaxy tend to be interleaved in radius and that every subsequent infall of the secondary happens from the other side than the previous one. In a real Type I shell galaxy, we know that the secondary had to arrive to the primary along the axis of the shell system, but we do not know from which side. Therefore there is the sign of the first generation, $o_I$, which can take the value of +1 or -1, depending on the direction of the first collision of the primary and the secondary and our definition of the positive and negative side of the observed shell system.

 However Dupraz \&~Combes \cite{dupcomb87}
did not try to account for the exact observed positions of shells.
Considering a three generation model, we showed
that the observed shell positions in NGC\,3923 can be reproduced in MOND so that the modeled shell radii differ by at most by $\sim$5\% from the observed values \cite{bil13}.

The multi-generation formation of a shell system was already revealed
in self-consistent simulations \cite{1990dig..book..216S,1994A&A...290..709S,katka11}.
The observations also support the multi-generation model: 
for some galaxies, the radial range of shells (the ratio of radii of the innermost to outermost shell) is so high that it cannot be produced in a single generation, as the simulations confirm \cite{DC86}.
Furthermore, the simulated one-generation shell systems never show as many shells as observed in some galaxies \cite{DC86}.

Type\,II and~III shell systems can also be, in principle, used to constrain the gravitational field of their host galaxies, but it is much more difficult. We know only about one attempt made by Kirihara et~al.~\cite{m31sh}. The authors run many   Newtonian simulations with dark matter to simulate the formation of shells and tidal features around M\,31. They found that the dark halo of the galaxy has  a steeper density falloff than expected from the cold dark matter cosmological simulations.

\section{Shell kinematics} \label{sec:kin}

Merrifield \&~Kuijken proposed a way to measure the gravitational potential of galaxies from the spectra of shells \cite{1998MNRAS.297.1292M}.
Using the approximation of a stationary shell, they derived that the line-of-sight
velocity distribution (LOSVD) near the shell edge should have double-peaked shape. However J\'{ i}lkov\'{ a} et~al.~\cite{2010ASPC..423..243J} and Ebrov\'{a} et~al.~\cite{2011efgt.book..225E} showed that the phase velocity of shells is substantial and therefore the LOSVD must rather take a quadruple-peaked shape. The shell velocity is typically tens of kilometers per second but could be more than 100\,km$/$s for giant ellipticals. 

\subsection{Nature of the quadruple-peaked shape}

\begin{figure*}
 \includegraphics[width=7.5cm]{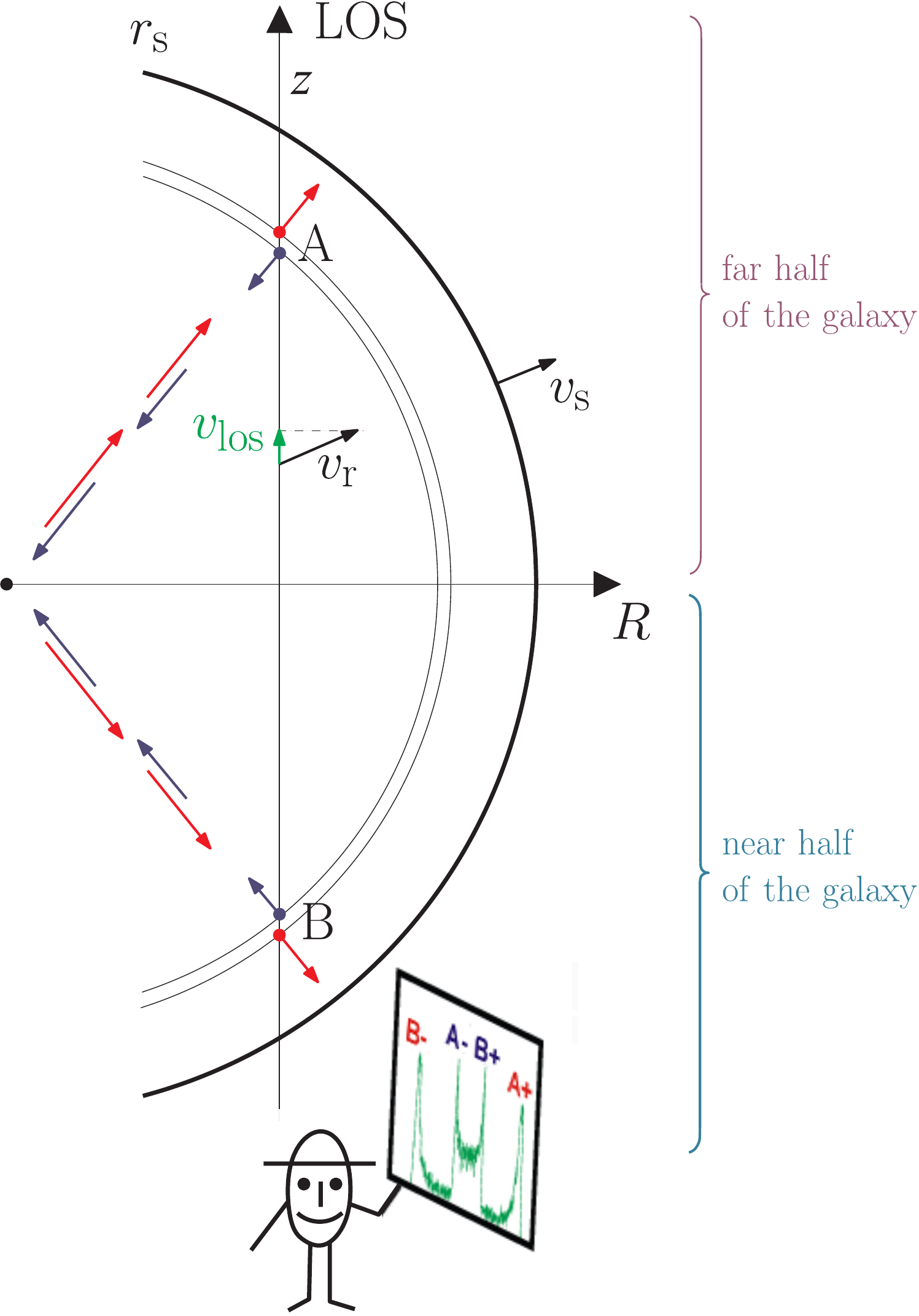} \includegraphics[width=7.5cm]{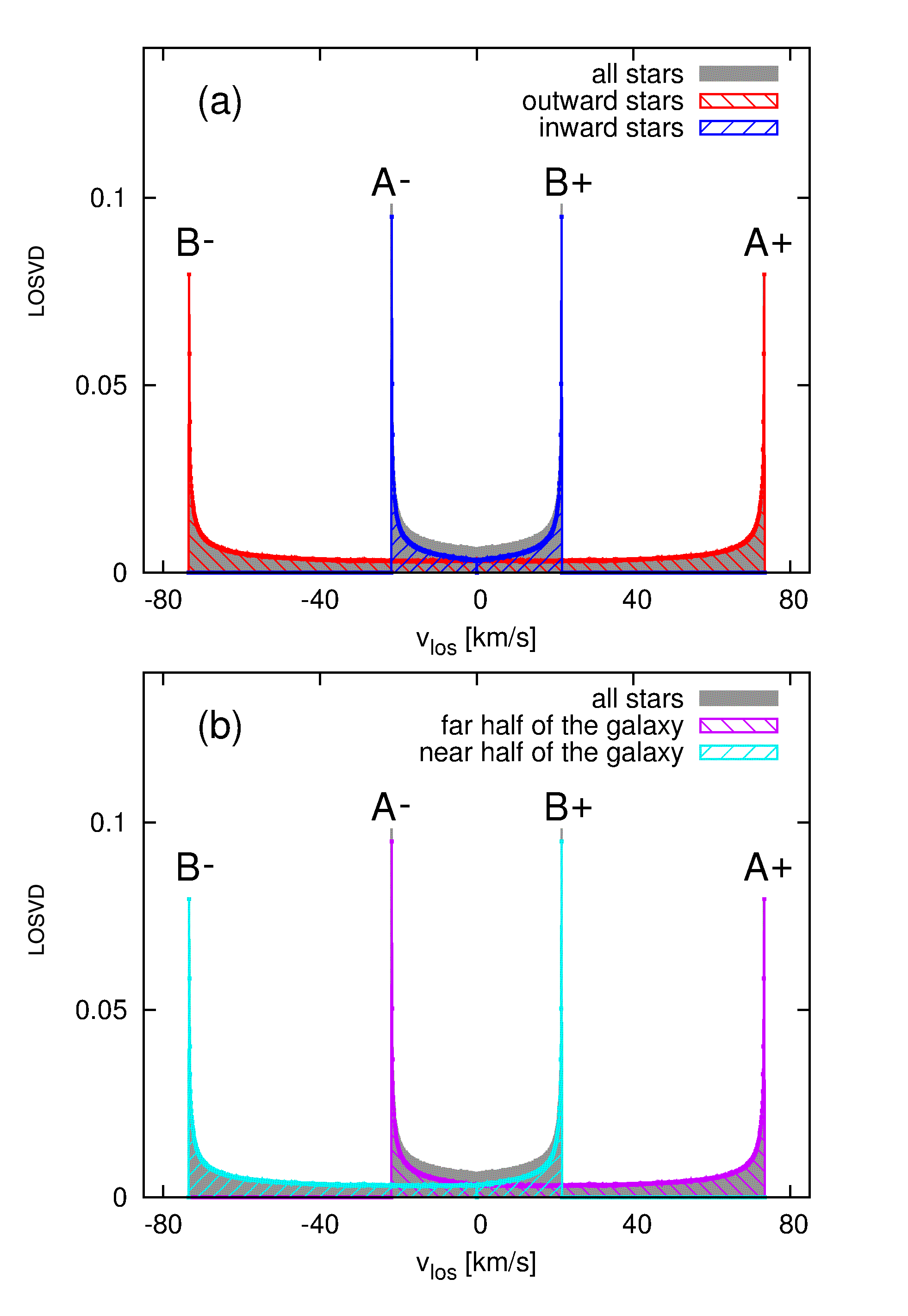}
\caption{%
Left: Scheme of the kinematics of a shell with radius $r_{\mathrm{s}}$
and phase velocity $v_{\mathrm{s}}$. The shell is composed of stars
on radial orbits with a radial velocity $v_{\mathrm{r}}$ and a LOS
velocity $v_{\mathrm{los}}$. Right: The LOSVD at the projected radius
$R=0.9r_{\mathrm{s}}$. The profile does not include stars of the
host galaxy, which are not part of the shell system. (a) The LOSVD
showing separate contributions from inward and outward stars; (b)
the same profile, separated for contributions from the near and far
half of the host galaxy. \label{fig:Mr.Eggy}
}
\end{figure*}

Let us approximate the shell system by the same simplified model as in Sect.~\ref{sec:rad}.
We will be interested only in the stars that get through the center of the galaxy exactly $n$ times (not counting the initial release), $n\geq 1$. These stars form the $n$-th shell.
The LOSVD consists of the projection of radial velocities
of stars to the line-of-sight (LOS) from different galactocentric
radii, see Fig.~\ref{fig:Mr.Eggy}. 
Due to the interplay between the projection effects and the dependence of the LOS velocities of the stars on the galactocentric radius, the contribution to the maximal
intensity of the LOSVD (the peak) comes from a specific radius.
This radius and the value of the projected star velocity is different for outward and inward stars. 
The situation is the same for the near and far half of the host
galaxy but with the sign of the star projected velocity revered (Fig.~\ref{fig:Mr.Eggy},
right). It results in a symmetric quadruple shape of the LOSVD.

\begin{figure}
\centering{}\includegraphics[width=12cm]{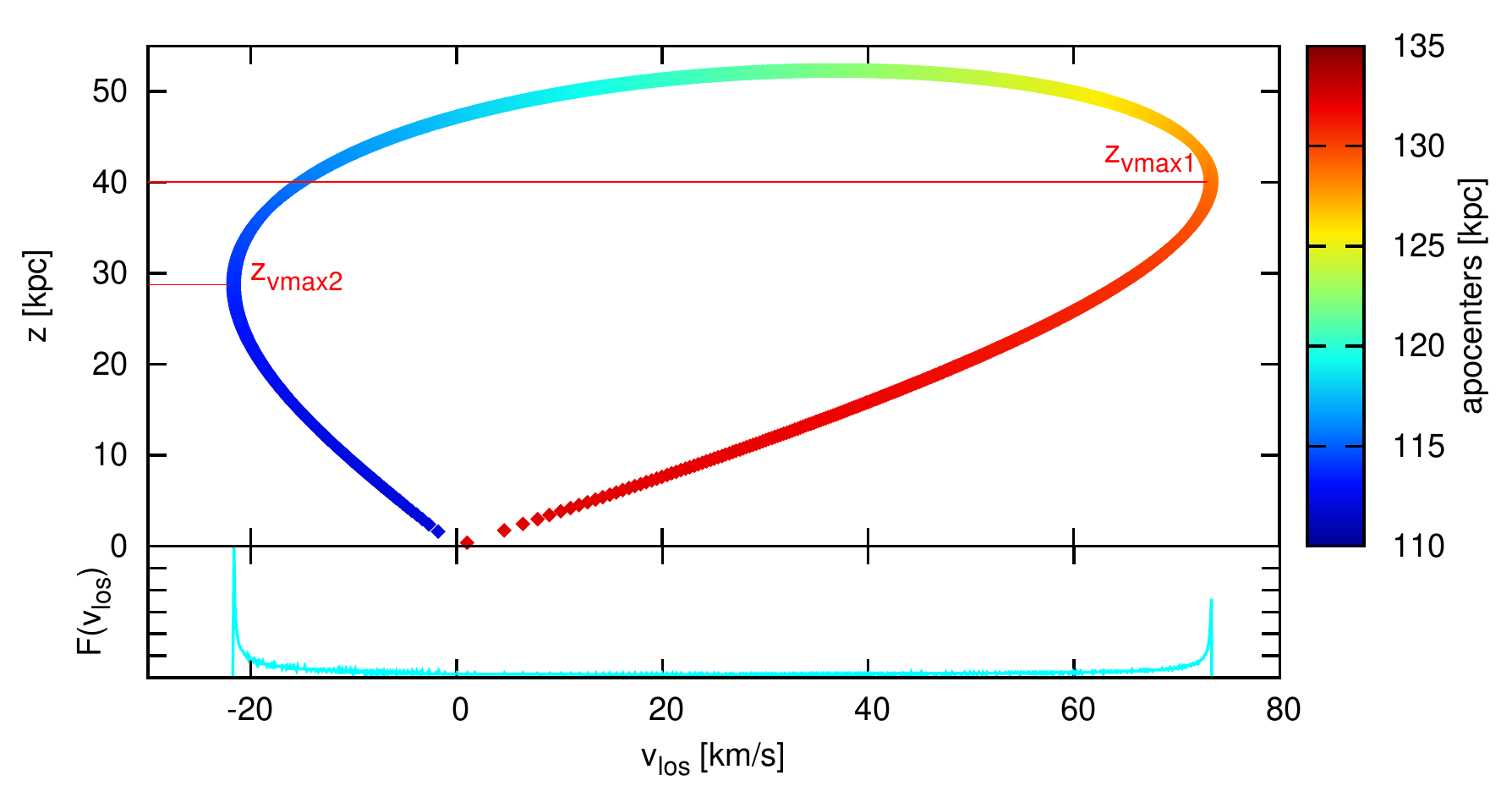} \caption{%
LOSVD and its individual contributions along the line-of-sight for
the far half of the host galaxy at the projected radius $R=0.9r_{\mathrm{s}}$.
\label{fig:barevny-apoc} 
}
\end{figure}

As can be seen in Fig.~\ref{fig:barevny-apoc}, contributions from
the far half of the host galaxy to the LOSVD form a curved line in
the space of the line-of-sight $z$ versus the line-of-sight velocity {$v_{\mathrm{los}}$}.
The colors encode the positions of the apocenters of the stars which
are currently at given $z$. The bottom panel of Fig.~\ref{fig:barevny-apoc}
shows the LOSVD itself. For the shape of the LOSVD, the dominant effect
is the bending of the curve in the $v_{\mathrm{los}}$ vs. $z$ plane around
two points, $z_{v\mathrm{max1}}$ and $z_{v\mathrm{max2}}$, at which
the maximal intensity of the LOSVD originates. At the same time, these
points naturally correspond to the LOS velocity extremes. The whole quadruple-peaked
shape is obtained by adding the contribution from the other half
of the galaxy.

\begin{figure}
\centering{}\includegraphics[width=0.6\textwidth]{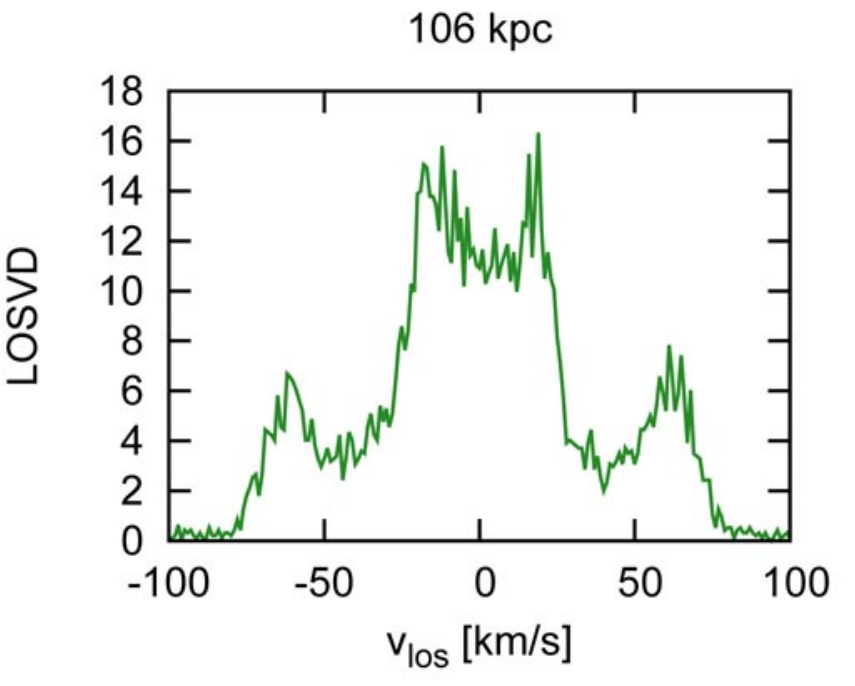} \caption{%
LOSVD (in arbitrary units) near the shell with radius of 112\,kpc from a self-consistent simulation of a radial minor merger with the total mass of the primary 8.2$\times10^{12}$\,M$_{\odot}$. Only particles originally belonging to the secondary are considered. The titles denote the projected radius of the respective LOSVD. 
\label{fig:skprof} 
}
\end{figure}

In the simplified model, we assumed a uniform distribution of shell stars in azimuth. In the real shell galaxies, this would not be satisfied. It would lead to a redistribution of the intensities of the four peaks but it should not change their positions. Moreover, real stars would not be released all at once in radial trajectories so, in reality, peaks would exhibit some degree of blurring. The blurring of the peaks can be seen in results of a self-consistent simulation of a radial merger in Fig.~\ref{fig:skprof}, but the peaks are still visible.

\subsection{Measuring gravitational acceleration from shell spectra}

We still assume the simplified model from Sect.~\ref{sec:rad}.
Positions of the four peaks of the LOSVD, at the particular time after
the merger, are determined by the potential of the host galaxy. 
Using the approximations described in Ebrov\'{a} et~al.~\cite{ebrova12}, we derived simple equations connecting the variables inherent to the potential of the host galaxy -- the circular velocity $v_{\mathrm{c}}$ at the shell edge radius and the current shell
phase velocity $v_{\mathrm{s}}$ -- with the observable quantities
\begin{equation}
v_{\mathrm{c}}=\frac{\left|v_{\mathrm{los,max}+}-v_{\mathrm{los,max}-}\right|}{2\sqrt{\left(1-R/r_{\mathrm{s}}\right)\left[1-4\left(R/r_{\mathrm{s}}\right)^{2}\left(1+R/r_{\mathrm{s}}\right)^{-2}\right]}},\label{eq:vc,obs}
\end{equation}
 
\begin{equation}
v_{\mathrm{s}}=\frac{v_{\mathrm{los,max}+}+v_{\mathrm{los,max}-}}{2\sqrt{1-4\left(R/r_{\mathrm{s}}\right)^{2}\left(1+R/r_{\mathrm{s}}\right)^{-2}}}.
\label{eq:vs,obs}
\end{equation}

The observable quantities are the radius of the LOSVD measurements $R$, the shell edge radius
$r_{\mathrm{s}}$, and the position of the LOSVD peaks $v_{\mathrm{los,max}+}$ and $v_{\mathrm{los,max}-}$. 
The position of the outer peak at one side of the LOSVD, $v_{\mathrm{los,max}+}$, and the position of the inner peak at the other side, $v_{\mathrm{los,max}-}$, correspond to the peaks A+ and A-, or B- and B+, respectively, in Fig.~\ref{fig:Mr.Eggy}. 
Similar relations were also derived by other authors \cite{fardal12,2013MNRAS.435..378S}.

The approximation, and thus Eqs.~(\ref{eq:vc,obs}) and (\ref{eq:vs,obs}), are the more accurate the closer to the shell edge the measurement of the LOSVD is performed, although they could be invalid immediately before the shell edge (see Ref.~\cite{ebrova12} for more details). The peaks of the LOSVD are the closer to each other the closer to the edge of the shell the measurements are done. Thus it is easer to resolve the peaks farther from the edge. But the surface brightness of the shell is highest near the edge. It is necessary to find a compromise for the projected radius of measurements.

The value of the circular velocity directly determines the gravitational acceleration at the radius of the shell edge.
On the other hand, the shell velocity depends on the shell number and on
the whole shape of the host galaxy potential from its center up to the
radius of the shell. Thus the interpretation of $v_{\mathrm{s}}$ is less straightforward but
it still carries a valuable information, see Sect.~\ref{sec:shvel}.

We verified the functionality of Eqs.~(\ref{eq:vc,obs}) and (\ref{eq:vs,obs})
on a result of test-particle simulation of a shell galaxy formation \cite{ebrova12}.
The typical deviation of the measured circular velocity was of the
order of units of km$/$s but the derived shell velocity was systematically
larger by 0--30\%. That can be caused by the non-radiality of the trajectories of
the stars constituting the shells or by a poor definition of the
shell radius in the simulation. We also applied Eqs.~(\ref{eq:vc,obs})
and (\ref{eq:vs,obs}) to the results of a self-consistent simulation
(yet unpublished). The maximal deviation of the measured circular
velocity was 15\% of the true value.

So far, the only convincing observation of shell kinematics was made by Fardal et~al.~\cite{fardal12}. They observed peaks in the LOSVD of red giant branch
stars in the region of the so-called Western Shelf in M31. 
They see rather asymmetric double-peaked structure since the ``stream-like shell'' is mostly located in the near half of the Andromeda galaxy.

\section{Shell velocities in MOND} 
\label{sec:shvel}
In the previous section we have learned that the lines in the shell spectra have quadruple-peaked profile and that the separation of its peaks allows to measure the circular velocity at the  shell edges (a quantity directly related to gravitational acceleration) and the phase velocity of the shell itself. In the paper~\cite{bil14b}, we focused on the expected values of these quantities in MOND. In MOND, unlike in the dark matter theory, these two velocities should take specific values. 

As expected from the basic tenets of MOND, the circular velocity asymptotically settles at the value of 
\begin{equation}
	V_\mathrm{c} = \sqrt[4]{GMa_0},
	\label{eq:vcirc}
\end{equation}
where $M$ denotes the baryonic mass of the galaxy. Therefore, if the asymptotic circular velocity is measured for a lot of shell galaxies, we expect that they will obey the same Tully--Fisher  (TF) relation, which is observed for disk galaxies. This behavior would be puzzling from the point of view of the dark matter theory, because the formation histories of elliptical and disk galaxies and their dark haloes are expected to be very different. 

The behavior of the shell phase velocity in MOND is not so straightforward to guess. We derived \cite{bil14b} that the velocity of the $n$-th shell asymptotically reaches the value of
\begin{equation}
	V_{\mathrm{s}, n} = \frac{V_\mathrm{c}}{\sqrt{2\pi}(n+1/2)}.
	\label{eq:vsh}
\end{equation}
This relation has several interesting implications. The shell number of a distant observed shell can be directly computed from Eq.~(\ref{eq:vsh}), once $V_\mathrm{c}$ and $V_{\mathrm{s}, n}$ are measured for it. If a lot of shell velocities were measured, not necessarily from the same galaxy, an analogy of the TF relation would hold
\begin{equation}
	\log V_{\mathrm{s}, n}  =\frac{1}{4}\log M-\log(n+1/2)+\frac{1}{4}\log\frac{Ga_0}{4\pi^2}.
	\label{eq:shtf}
\end{equation}
This relation has several branches for different shell numbers $n$. The separation between the individual branches decreases with increasing $n$.  When the velocities $v_\mathrm{c}$ and $v_\mathrm{s}$ are measured for a lot of shells, we can see from Eq.~(\ref{eq:vsh}), that the histogram of $v_{\mathrm{c}}/v_{\mathrm{s}}$ will form a series of equidistant peaks located at the  values of $\sqrt{2\pi}(n+1/2)$. We must remember that Eq.~(\ref{eq:vsh}) gives only the asymptotic value of the shell velocity, so that it is applicable for shells well beyond the MOND transitional radius $R_{\mathrm{T}} = \sqrt{\frac{GM}{a_0}}$  and the characteristic radius of the galaxy. However, MOND gravitational field in a galaxy is determined by the distribution of the baryonic matter. With the knowledge of the gravitational field we can calculate the expected velocity of the shell using Eq.~(\ref{eq:dpdr}).

\section{Shell identification}
\label{sec:shid}

Shells offer two ways to constrain the gravitational field in their host galaxies. Besides the shell spectra, the other uses the connection of shell radii and the free-fall-time. The first attempts to constrain the potential from the shell radii of Type\,I shell galaxies were made in 80's \cite{quinn84,DC86,hq87}. Methods of that time were based on the assumptions that 1) all the shells present in the galaxy are made of stars released at once from the center of the galaxy (the single-generation model) and 2) if the shell numbers $n$ and $m$ are observed, then also all the shells with numbers between $n$ and $m$ are observed (there are no ``missing shells'' escaping observations). 
If one of these assumptions does not hold in the observed galaxy, the method can lead to wrong conclusions.
One of the first results of the shell radii analysis was that the positions of shells in NGC\,3923 contradict MOND \cite{hqmond}. Shortly after that, new shells were discovered \cite{prieur88}. Besides this broken assumption, Milgrom found that the used method was applied incorrectly \cite{milgsh} . Moreover, then Dupraz~\&~Combes  came with the multi-generation formation of shell galaxies \cite{dupcomb87} (Sect. \ref{sec:rad}). The galaxy NGC\,3923 is a  more probable candidate on the multi-generation origin than any other shell galaxy due to its record shell radial range (see Sect.~\ref{sec:rad}), which is 60. 
Due to the advent of the multi-generation scenario, the interest in shells as probes of gravitational potential faded. 

The next attempt was the method of shell identification by B\'{i}lek et~al. presented in Ref.~\cite{bil13} and further developed in Ref.~\cite{bil14a}.
This method circumvents the complications mentioned above. We call this method the shell identification because it tries to reveal the ``identity'' of the observed shells -- their shell numbers and the generations they come from.
It allows to test the compatibility of the observed shell radial distribution with the tested gravitational potential. The method is well suited for testing of the MOND gravitational theories, because gravitational field in these theories is unambiguously given by the distribution of the visible matter. Moreover, the morphology of the shell system constrains the spatial matter distribution of its host galaxy \cite{DC86}. 
The method is based on the assumption that we are able to predict the time evolution of shell radii in the studied galaxy. 
It is questionable how well our approximation \eqref{eq:pos2} works in a real galaxy. Anyway, the method stays applicable for any model of the shell radii evolution.

We will try to explain the principle of the method here. Suppose that we are able to describe the time evolution of shell radii in a given galaxy absolutely precisely and that we know that the shells were formed in a single generation. Then we could determine the shell numbers (and hence the original direction of the collision) very easily. We would simply calculate the time evolution of shell radii and find the time, at which the observed shell radii coincide with those modeled (and would even obtain the age of the shell system). We would only need to find the direction of the first arrival of the secondary. 
There are only two possibilities -- from one side of the symmetry axis of the shell system or from the other. If some shells were missed by observation, it would not destabilize the method, but, on the contrary, the method would hint us that the shells exist and what are the radii where they should be found.
 
The situation becomes a little more complicated if the system was formed in two generations. Then we would have to find the direction of the first collision and two times $t_1$ and $t_2$, so that one part of the observed shells is reproduced by the model for the time $t_1$ and the remaining part for $t_2$. 
The direction of the second passage has to be opposite to that of the first one. If the shell system was created in more than two generations, a new condition  appears: the age difference between subsequent generations shortens, as the amplitude of the oscillations of the secondary is damped by dynamical friction. 

Further, suppose that we want to test the presence of a given potential in an observed galaxy and we are able to describe the time evolution of shell radii absolutely precisely and we know the number of generations in the galaxy. If the potential is correct, then it is possible to identify the shells meeting the criteria from the previous paragraph. If it is not possible, then the potential is incorrect. 

For the real situation, things complicate even more. We cannot describe the shell radius evolution absolutely precisely as well as we cannot measure the radii of the observed shells exactly.
We must replace the condition of the equality of the modeled and observed shell radii by the condition of ``approximate equality''. But what does it precisely mean, that modeled and observed shell radii are ``approximately equal''? Nobody knows at the moment. This is the biggest weakness of the method. However, this criterion is obtainable from self-consistent simulations. The simulations will have to include different kinds of structure of the secondary, non-radial mergers, ellipticities of the primary, etc. So far, we have compared the analytical approximation given by Eq.~(\ref{eq:pos2}) with one Newtonian self-consistent simulation with dark matter \cite{bil14a}. The deviations of the shell radii in the simulation from the analytic approximation were lower than 4\%.

The second unknown factor, for a real galaxy, is the total number of generations. Again, this is hard to guess at the moment. In the small number of Newtonian self-consistent simulations that have been performed, maximally 4 generations of shells were formed \cite{1994A&A...290..709S,katka11, cooper11}.
The amount of generations is probably related to the difference of the potential energy at the positions of the innermost and outermost shells of the system.

The third and last open question is the amount of missing shells. If a lot of shells is allowed to be missing, almost any set of shell radii can be consistent with the given potential. This number can be probably constrained from the statistics of surface brightness of shells, which nobody has done so far. 

\subsection{Consistency of MOND with the shell radial distribution in NGC\,3923}

\begin{figure}[tb]
\centering{}\includegraphics[width=0.6\textwidth]{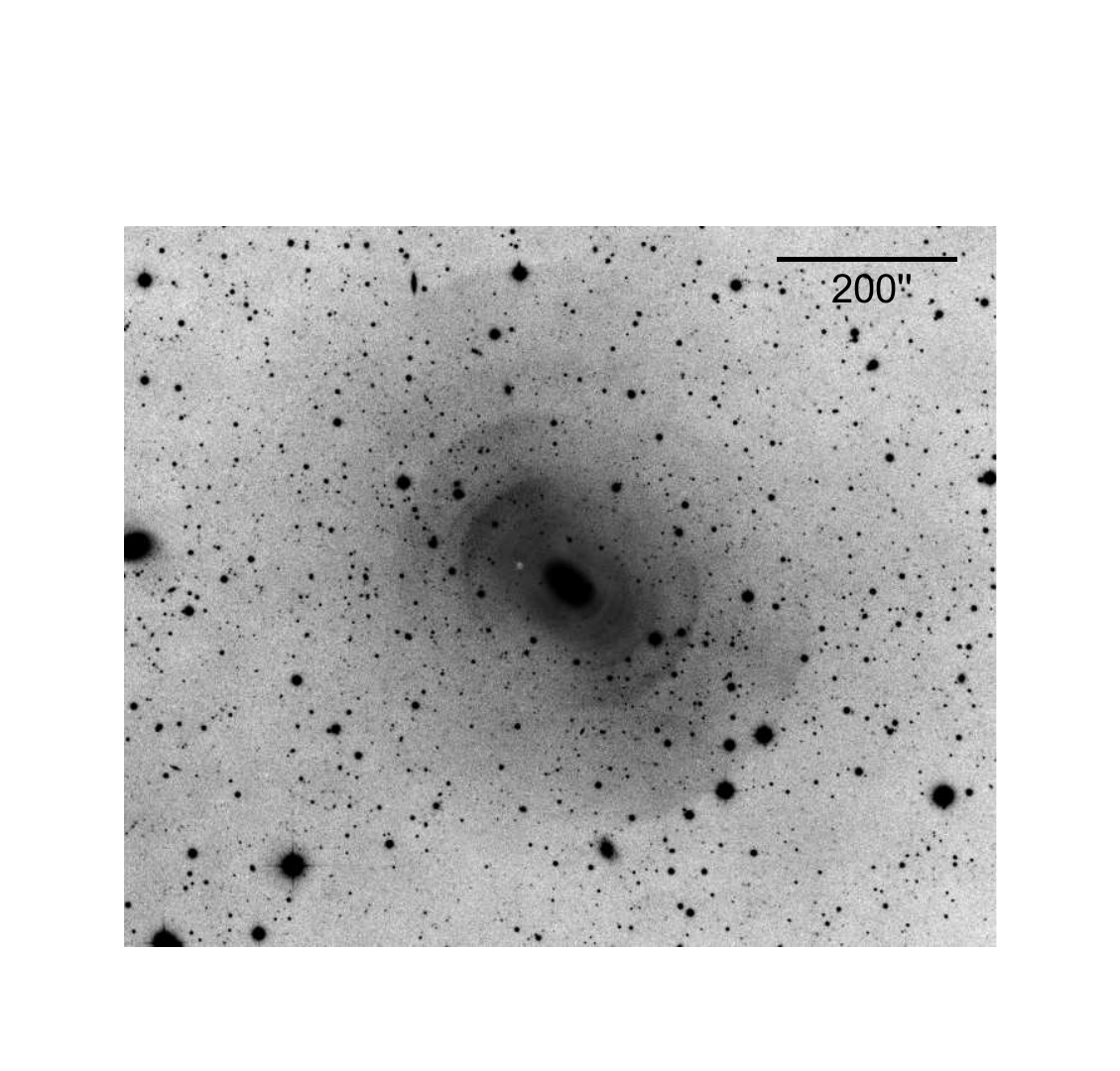} 
\vspace{-6ex}
\caption{%
Type\,I shell galaxy NGC 3923. Courtesy of David Malin, Australian Astronomical Observatory.}
\label{fig:N3923} 
\end{figure}

\begin{table}[tb]
\tiny
\centering
\begin{tabular}{crrr}
\hline\hline
Label & $r$ $[\arcsec]$ & $d$ $[$kpc$]$ & $a$ $[a_0]$ \\
\hline
a  &	$+1170$ &	$130.7$ &	$0.2$ \\
b  &	$-840$ &	$93.8$ &	$0.3$ \\
c  &	$+630$ &	$70.4$ &	$0.4$ \\
d  &	$-520$ &	$58.1$ &	$0.4$ \\
e  &	$+365$ &	$40.8$ &	$0.6$ \\
f  &	$-280$ &	$31.3$ &	$0.8$ \\
g  &	$+203$ &	$22.7$ &	$1.1$ \\
h  &	$-148.5$ &	$16.6$ &	$1.5$ \\
i  &	$+147.3$ &	$16.5$ &	$1.5$ \\
j  &	$+128.1$ &	$14.3$ &	$1.7$ \\
k  &	$-103.6$ &	$11.6$ &	$2.2$ \\
l  &	$+99.9$ &	$11.2$ &	$2.2$ \\
m  &	$-79.6$ &	$8.9$ &	$2.9$ \\
n  &	$+72.8$ &	$8.1$ &	$3.1$ \\
o  &	$-67.0$ &	$7.5$ &	$3.4$ \\
p  &	$+64.1$ &	$7.2$ &	$3.6$ \\
q  &	$+60.4$ &	$6.7$ &	$3.9$ \\
r  &	$-55.5$ &	$6.2$ &	$4.2$ \\
s  &	$+51.2$ &	$5.7$ &	$4.6$ \\
t  &	$-44.0$ &	$4.9$ &	$5.5$ \\
u  &	$+41.5$ &	$4.6$ &	$5.9$ \\
v  &	$-37.7$ &	$4.2$ &	$6.6$ \\
w  &	$+34.3$ &	$3.8$ &	$7.3$ \\
x  &	$+29.3$ &	$3.3$ &	$8.8$ \\
y  &	$-28.7$ &	$3.2$ &	$9.0$ \\
A  &	$+19.4$ &	$2.2$ &	$14$ \\
B  &	$-18.0$ &	$2.0$ &	$15$ \\

\end{tabular}
\label{tab:sh}
\caption{Shells of NGC\,3923. $r$\,--\,the observed distance of the shell from the center of NGC\,3923, data compilation taken from Ref.~\cite{bil13}. The plus sign means that the shell is situated on the north-eastern side of the galaxy and minus on the south-western; $d$\,--\,the radius of the shell in kiloparsecs assuming the galaxy distance of 23\,Mpc; $a$\,--\,the gravitational acceleration at the radius of the shell for the potential used in Ref.~\cite{bil13}.}
\label{tab:shpos}
\end{table}

Although these problems may look serious, the method of shell identification is able to give meaningful answers when we ask the right questions. 
We have used the method in two papers so far, which we are going to review here. 
We tested the compatibility of MOND with the radial distribution of shells in a famous, well studied shell galaxy NGC\,3923 (Fig.~\ref{fig:N3923}) \cite{bil13}. It is a relatively nearby ($\sim$23\,Mpc) elliptical with many record properties. Besides the mentioned extraordinary shell radial range, it also hosts the biggest known shell with the radius of $\sim$120\,kpc. It is also the galaxy with the highest known number of shells, namely 27 (Table~\ref{tab:shpos}). The galaxy was imaged in infrared by the Spitzer telescope. 
We derived its mass profile on the basis of the image in the 3.6\,$\mu$m band from this telescope. The mass-to-light ratio is relatively constant in this band \cite{constml}. 
Moreover, to constrain the mass-to-light ratio even more, we used its observed correlation with the luminosity in $J$ and $K_s$ bands. 
Since the minor axis rotation is observed in NGC\,3923 \cite{minrot}, we treated the galaxy as a prolate ellipsoid with the major axis oriented perpendicularly to the line-of-sight. 
Now, we were able to calculate the MONDian gravitational potential along the major axis of the galaxy with the use of the algebraic relation $a\mu(a/a_0) = a_N$ between the Newtonian acceleration $a_N$ and MONDian acceleration $a$.
When we learned the potential and assumed that the secondary is stripped only when passing through the center of the primary, the time evolution could be calculated using Eq.~(\ref{eq:pos2}).

\begin{figure}
\centering{}\includegraphics[width=\textwidth]{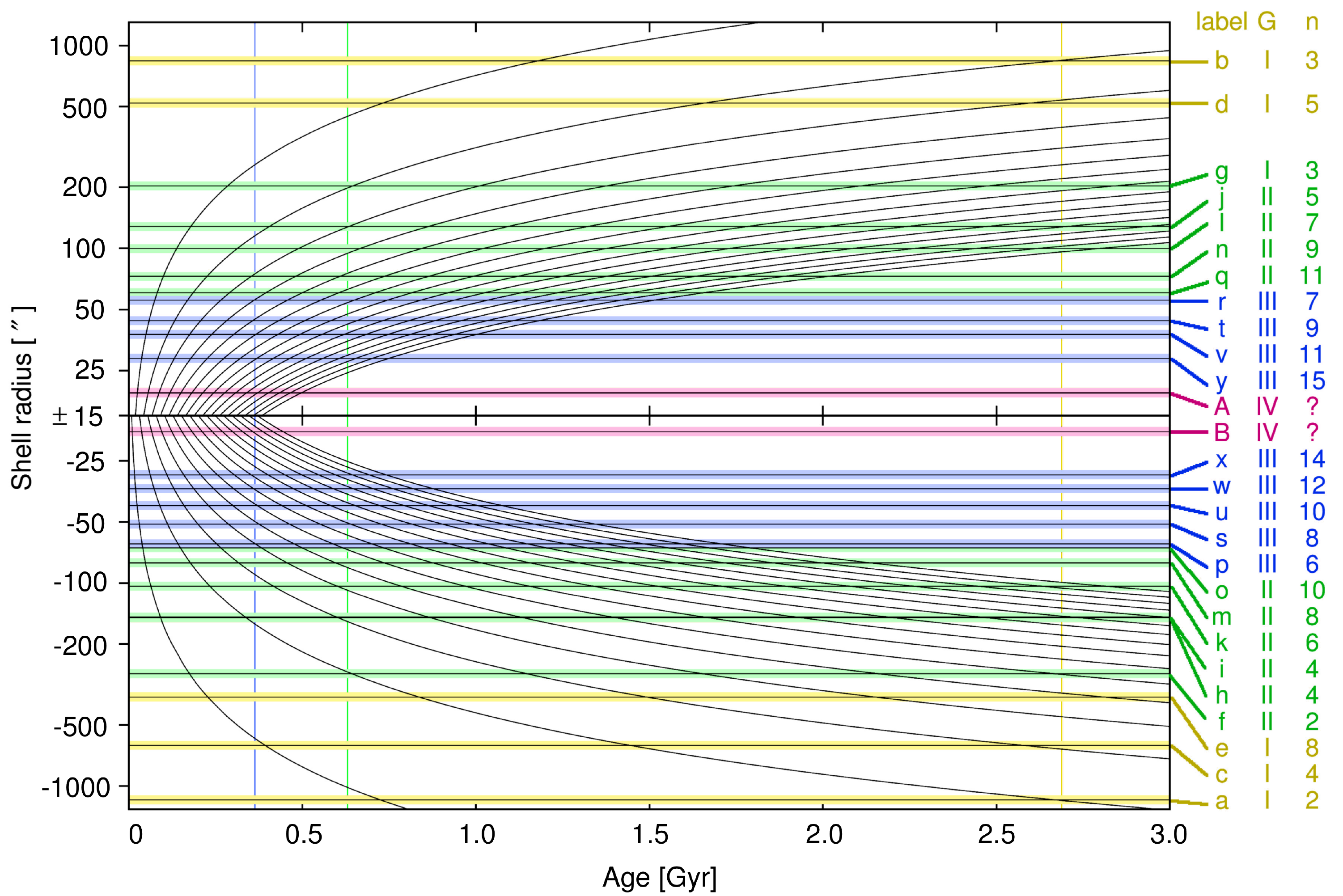} \caption{%
Calculated time evolution of shell radii in the potential of NGC\,3923 used in Ref.~\cite{bil13} -- black curves. The observed radii of the shells of NGC\,3923 plus 5\% uncertainty boundaries -- horizontal lines. The shells identified to come from the same generations are marked by identical colors. The age of a generation is marked by the vertical line of the same color. The calculated shell radii should ideally reach their observed values at the times marked by the vertical lines. The observed shells are named by their labels. The other two columns on the right denote the identified generation to which the shell belongs and its shell number. The positivity/negativity of the radius of an observed shell in this figure is determined by its identified shell number, the generation, and by the side at which the shell lies in the galaxy. The sign of the shell $i$ was switched since it is interpreted as the shell $h$ encircling the galaxy. The two innermost shells probably come from the fourth generation and were not formed in the manner that our model assumes.}
\label{fig:shcomp} 
\end{figure}

It came out that the radii of the observed 25 outermost shell of NGC\,3923  can be identified in a very satisfactory manner (Fig.~\ref{fig:shcomp}). The modeled shells deviated from those observed maximally by 5.4\%. This is not a bad result considering that we used one fixed potential derived from quantities independent of the shell radii. 
We cannot expect any significantly better coincidence, because the observational precision of the radii of some shells is comparable to this value. 
We needed three generations to explain the shell distribution. 

The radii of the innermost two shells are not explained well. Some of the assumptions of our model of shell propagation was possibly violated. For example, in the simulation of Cooper et~al.~\cite{cooper11} we can see that the secondary goes through the center of the primary three times, giving rise to three shell generations. However, the secondary does not dissolve at the center, but decays later at a certain non-zero galactocentric radius. From its last debris, a further, fourth, shell generation forms. This could also be the case of NGC\,3923. The evolution of the inner shells could be also affected by a slight non-radiality of the merger. 

The identification, we found, requires to postulate the existence of three missing shells. 
The first is escaping observations probably due to the dust cloud which is observed at its expected position. The explanation of the absence of the other two shells is not so easy as for the previous one. The azimuth of the brightest spot is not common for all the shells. Similar phenomenon probably occurs also in the direction parallel to the line-of-sight. Therefore the contrast of some shell edges could be lowered. 

Our identification requires that two of the reported shells are in fact one shell encircling the galaxy. This is supported by the facts that their radii differ only by 1\% and that one of them is much fainter than the other shells in its vicinity. A~similar effect sometimes happens in our test-particle simulations (see, e.g., \cite{ebrova12}) with spherical potentials, but when it happens, most shells encircle the galaxy. This may be connected with the ellipticity of the galaxy and the shell focusing effect \cite{hq89, DC86}. Only very few simulations of shell galaxies have been performed with elliptical primaries so far.

The found identification also implies many details about the history of the merger. The secondary impacted to the center of NGC\,3923 around 2.7\,Gyr ago. It reached the galaxy from southwests. Then its peeled remainder was trapped into the potential well of NGC\,3923 and started to oscillate. It went again through the center 650 and 350\,Myr ago and got completely dissolved shortly afterward. 

\subsection{A MOND prediction of a new shell in NGC\,3923}
\begin{figure*}
\resizebox{0.9\hsize}{!}{\includegraphics{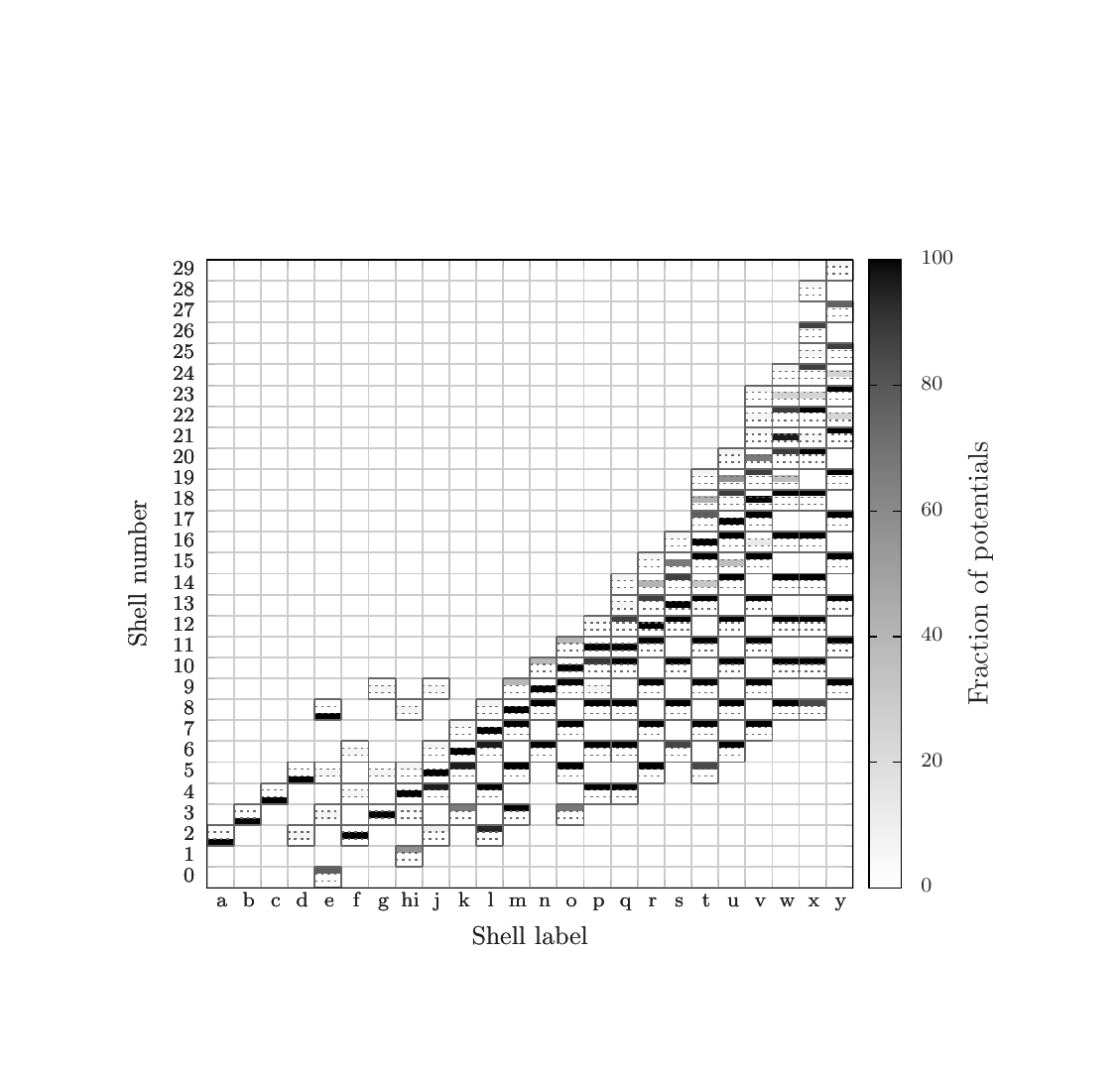}}
\caption{All satisfactory identifications found in the basic grid of gravitational potentials defined in Ref.~\cite{bil14a}. An identification of a~shell is the assignment of a~shell number and shell generation to it. The sub-cells of every cell denote the identified generation of the shell -- bottom part: the first generation; middle part: the second generation; upper part: the third generation. The color of the sub-cell encodes the number of the potentials from the basic grid for which the shell was successfully identified in the given way, divided by the total number of potential in the basic grid. We can see that the three outermost shells ($a$, $b$ and $c$, see Table~\ref{tab:shpos}) must be unambiguously identified as the shells number 2, 3, 4 from the first generation and for all the potentials the fourth outermost shell $d$ can be identified as the shell number 5 from the same generation. These findings imply that the shell number 1 must exist.}
\label{fig:idtab_basgrid}
\end{figure*}

In our paper \cite{bil13}, the identification was found ''by eye``. Because this procedure is slow and subjective, we developed a code for the shell identification. The code takes as the input the set of observed shell radii and a potential to be tested. It finds all identifications of the observed shells in the potential meeting our identification criteria. For the reasons described in the paper \cite{bil13}, we defined the acceptable identification so that the observed shell radii do not deviate more than by 7\% from those modeled, the system was created in maximally 3 generations, and 2 shells are allowed to be missing in every generation. If no acceptable identification can be found, the potential is pronounced inconsistent with the observed shell distribution. The code was applied to around 15000 more or less probable MOND potentials of NGC\,3923. They include various values of the mass-to-light ratio, distance of the galaxy from the Earth, values of the acceleration constant $a_0$ and modified mass profiles. The immediate result was that the identification of the galaxy's shells becomes the less unique the closer to the galactic center the shell lies (Fig.~\ref{fig:idtab_basgrid}). There are often five or more ways to assign the shell and generation numbers to a given observed shell. This is not a fault of MOND or of our method. The explored gravitational potentials had the property that they create densely spaced shells near their center. When this is combined with the benevolently chosen tolerance of difference between the observed and modeled shell radii, the identification must become ambiguous. 

We therefore focused on the properties that are common for all identifications. For any potential compatible with the shell distribution of NGC\,3923, at least one of the found identifications (but, in fact, a vast majority of them) assigns the shell numbers 2, 3, 4, 5 to the four outermost shells and classifies them into the first generation (Fig.~\ref{fig:idtab_basgrid}). This fact has an interesting implication -- the shell number 1 must exist. It should be located at the distance of around 1900$^{\prime\prime}$ on the southwestern side of the galaxy. But its surface brightness can be very low. Shells end their life by growing so big, that only very few stars can make them up. If the predicted shell is constituted by the same number of stars as the currently biggest one at 1170$^{\prime\prime}$, its surface magnitude should be around 28\,mag/arcsec$^2$ in the B band, which is in the reach of the contemporary instruments. For all the found identifications, three generations are necessary to explain the observed shell distribution. In all of them, the secondary had to hit NGC\,3923 from the southwestern side at their first encounter.

To test the ability of the shell identification method to predict positions of new shells, we excluded the known shell at 1170$^{\prime\prime}$ from the sample and tried to reconstruct its position on the basis of the positions of the remaining shells repeating the same analysis. The reconstructed value was almost precise: 1160$^{\prime\prime}$.

\section{Summary}
\label{sec:sum}
We have seen that the shells appear in about 10\% of all elliptical galaxies, but this number is environmentally dependent. About one third of shell galaxies is of the Type\,I, which is the best suitable type of shell galaxies for constraining the gravitational potential. If we are able to model shell radii in a given potential, we can test for  presence of a given potential in the shell galaxy using the shell identification method. In this method we check, whether the observed shell radial distribution is compatible with the modeled radii. The shell identification method allowed to verify the compatibility of MOND with the observed shell  distribution in a famous shell elliptical galaxy NGC\,3923 \cite{bil13}. The method can be also used to predict the position of yet undiscovered shells. Applying the method of shell identification to a large number of possible MOND potentials of NGC\,3923, we concluded that a new, yet undiscovered and the biggest shell of NGC\,3923 must exist at the distance of 1900$^{\prime\prime}$ south-west from the galactic center \cite{bil14a}. The spectra of shells allow to measure the circular speed at  the radius of the shell edge and the phase velocity of the shell \cite{ebrova12}. In MOND, these velocities are predicted to asymptotically converge to the values given by Eq.~(\ref{eq:vcirc}) and Eq.~(\ref{eq:vsh}) \cite{bil14b}, respectively, similarly as rotational curves of disk galaxies get flat at large radii. These facts imply that the circular velocity in elliptical galaxies should obey the Tully--Fisher relation known for disk galaxies and the shell velocity forms a multi-branched analogy of the Tully-Fisher relation. 

\section*{Acknowledgements}
We express our thanks to  I. Orlitov\'{a} and L. Janekov\'{a} for useful comments. The work of IE was partially supported from the EU grant GLORIA (FP-7 Capacities; No.~283783).
We acknowledge the support from the following sources: Czech support for the
long-term development of the research institution RVO67985815
(MB, IE, BJ, and KB), the project SVV-260089 by the Charles University
in Prague (MB, and IE) and the grant MUNI/A/0773/2013 by the Masaryk
University in Brno (KB). 
Access to computing and storage facilities owned by parties and projects
contributing to the National Grid Infrastructure MetaCentrum, provided
under the programme "Projects of Large Infrastructure for Research,
Development, and Innovations" (LM2010005), is greatly appreciated.
Access to the CERIT-SC computing and storage facilities provided under
the programme Center CERIT Scientific Cloud, part of the Operational
Program Research and Development for Innovations, reg. no. CZ.
1.05/3.2.00/08.0144, is greatly appreciated.

\let\OLDthebibliography\thebibliography
\renewcommand\thebibliography[1]{
  \OLDthebibliography{#1}
  \setlength{\parskip}{0pt}
  \setlength{\itemsep}{0.5ex}
}
\bibliographystyle{bibgen}
\bibliography{citace}

\end{document}